\documentclass[12pt,a4paper]{article}
\usepackage[latin1]{inputenc}
\usepackage[english]{babel}
\usepackage{graphicx}
\def\la{\mathrel{\mathchoice
{\vcenter{\offinterlineskip\halign{\hfil
$\displaystyle##$\hfil\cr<\cr\sim\cr}}}
{\vcenter{\offinterlineskip\halign{\hfil$\textstyle##$\hfil\cr
<\cr\sim\cr}}}
{\vcenter{\offinterlineskip\halign{\hfil$\scriptstyle##$\hfil\cr
<\cr\sim\cr}}}
{\vcenter{\offinterlineskip\halign{\hfil$\scriptscriptstyle##$\hfil\cr
<\cr\sim\cr}}}}}

\begin{document}
\addtocounter{footnote}{1}
\title{Study of Stability of a Charged Topological Soliton in the System of Two Interacting Scalar Fields}

\addtocounter{footnote}{0}

\author{V. A. Gani$^*$,
N. B. Konyukhova$^{**}$,
S. V. Kurochkin$^{**}$,
and V. A. Lensky$^*$\\
\\
($^*$
\it\small Institute for Theoretical and Experimental Physics, Russian Federation State Scientific Center,\\
\it\small ul. Bol'shaya Cheremushkinskaya 25, Moscow, 117259 Russia;\\
$^{**}$
\it\small Dorodnitsyn Computing Center, Russian Academy of Sciences,\\
\it\small ul. Vavilova 40, Moscow, 119991 Russia)\\
}
\date{}
\maketitle

\begin{abstract}
An analytical-numerical analysis of the singular self-adjoint spectral problem
for a system of three linear ordinary second-order differential equations defined
on the entire real exis is presented. This problem comes to existence in the
nonlinear field theory. The dependence of the differential equations on the
spectral parameter is nonlinear, which results in a quadratic operator Hermitian
pencil.
\end{abstract}

\section{Introduction. Exact solution to a system of two nonlinear wave equations}

The construction of precise regular solutions in systems of interacting classical
fields and study of their dynamic stability are of great interest in modern nonlinear
field theory \cite{1}.

In this paper, we study the problem of stability of such a solution for a system
of two interacting scalar fields (this solution was reported in paper \cite{2}),
the neutral Higgs field and a charged linear field (the model was suggested
in \cite{FLS}). In the (1+1)-dimensional Minkowski space, the considered field
system is described by the Lagrangian
\begin{equation}
{\cal L}=|\partial_t\xi |^2-
|\partial_x\xi|^2+(\partial_t\phi)^2/2-
(\partial_x\phi)^2/2-h^2\phi^2|\xi|^2-m^2(\phi^2-v^2)^2/2.
\label{lagr}
\end{equation}
Here, $\phi$ is a real scalar field; $\xi$ is a complex scalar field; and $h$, $m$
and $v$ are real positive constants. We use the system of units where $c=\hbar=1$;
$c$ is the speed of light in vacuum and $\hbar$ is the Planck constant. In this
system of units, only the dimension of mass $M$ is nontrivial; the dimension of
length and time is $1/M$ (\cite{1}, p.~13).
In (\ref{lagr}) $\phi$, $\xi$, and $v$ are dimensionless quantities, and $[m]=[h]=M$.
It is convenient to turn to the dimensionless independent variables
\begin{equation}
\tilde{x}=(hv/\sqrt{2})\:x, \quad
\tilde{t}=(hv/\sqrt{2})\:t
\label{newvars}
\end{equation}
and introduce the new functions
\begin{equation}
\tilde{\phi}=\phi/v, \quad
\tilde{\xi}=\xi/v.
\label{newfields}
\end{equation}
In what follows, we use variables (\ref{newvars}) and functions (\ref{newfields}) and omit the tilde over the letters.

The system of the Lagrange-Euler equations for the Lagrangian (\ref{lagr})
in terms of (\ref{newvars}) and (\ref{newfields}) takes the form
\begin{equation}
\frac{\partial^2\xi}{\partial t^2}-\frac{\partial^2\xi}{\partial
x^2}+2\phi^2\xi=0,
\label{eulereq1}
\end{equation}
\begin{equation}
\frac{\partial^2\phi}{\partial t^2}-\frac{\partial^2\phi}{\partial
x^2}+4\xi\xi^*\phi+\frac{4}{\kappa^2}(\phi^2-1)\phi=0, \quad t, x \in {\bf R},
\label{eulereq2}
\end{equation}
where $\kappa$ is a positive dimensionless parameter and $\kappa^2=h^2/m^2$.
(Here and in what follows, the asterisk denotes the Hermitian conjugation.)
This system is invariant with respect to global (not depending on $x$ and $t$)
transformations $\xi\to\xi\exp(i\alpha)$ and $\phi\to-\phi$; for the motion integrals,
it has the energy integral $E$, charge $Q$, and topological charge $P$ defined by
the formulas
\begin{equation}
E=\int\limits_{-\infty}^\infty
\left[
\left|\frac{\partial\xi}{\partial t}\right|^2
+
\left|\frac{\partial\xi}{\partial x}\right|^2
+
\frac{1}{2} \left(\frac{\partial\phi}{\partial t}\right)^2
+
\frac{1}{2} \left(\frac{\partial\phi}{\partial x}\right)^2
+
2\phi^2|\xi|^2
+
\frac{1}{\kappa ^2}(\phi ^2 -1)^2
\right]dx,
\label{energy}
\end{equation}
\begin{equation}
Q = -i\int\limits_{-\infty}^\infty
\left(
\xi ^*\frac{\partial\xi}{\partial t}
-
\xi \frac{\partial\xi^*}{\partial t}
\right)dx,
\label{charge}
\end{equation}
\begin{equation}
P=\frac{1}{2}\int\limits_{-\infty}^\infty
\frac{\partial\phi}{\partial x}\:dx,
\label{tcharge}
\end{equation}
Thus, the quantities $E$ and $Q$ do not depend on time for those solutions
of system (\ref{eulereq1}), (\ref{eulereq2}), for which the integrals on the
right-hand sides of (\ref{energy}) and (\ref{charge}) converge, and $P$ (formula (\ref{tcharge}))
does not depend on $t$ for any $\phi(t,x)$ that has finite limits as $|x|\to\infty$ $\forall\: t\in\bf{R}$.

By definition, the invariance of Eqs. (\ref{eulereq1}), (\ref{eulereq2}) with respect to
the transformation $\xi\to\xi\exp(i\alpha)$, where $\alpha$ is an arbitrary real number,
implies global U(1)-symmetry of these equations \cite{1}, and charge (\ref{charge})
is called a U(1)-charge.

First of all, we note that system (\ref{eulereq1}), (\ref{eulereq2}) is known to have the following
particular solutions:\\
(1) the trivial solution $\xi\equiv 0$, $\phi\equiv 0$, which is called a false vacuum
since it has a nonzero energy density; traveling wave solutions over the false vacuum
$\xi=\psi(x\pm t)$, $\phi\equiv 0$, where $\psi$ is an arbitrary twice continuously
differentiable function (these solutions also have a nonzero energy density);\\
(2) solutions
\begin{equation}
\xi\equiv 0, \quad \phi_{\pm} = \pm 1
\label{truevacs}
\end{equation}
with zero energy, which are reffered to as true vacua; and\\
(3) solutions of the domain wall type
\begin{equation}
\xi \equiv 0 ,\quad \phi _{w}(x)=\tanh(\sqrt {2}\:x/\kappa),
\label{wall}
\end{equation}
that have the finite energy
\begin{equation}
E_{w}=4\sqrt{2}/(3\kappa),
\label{Ew}
\end{equation}
zero charge $Q$, and the nonzero topological charge
\begin{eqnarray}
P_{w}=1
\nonumber
\end{eqnarray}
(of course, the antiwall $\xi\equiv 0$, $\tilde{\phi}_{w}=-\phi_{w}$ is also a solution to
(\ref{eulereq1}), (\ref{eulereq2}) with the same energy (\ref{Ew}) and topological charge
$\tilde{P}_{w}=-1$).

{\bf Definition 1.} A {\it topological soliton} (or a {\it domain wall}) for system
(\ref{eulereq1}), (\ref{eulereq2}) is a solution existing and bounded in the entire
space and satisfying the conditions
\begin{eqnarray}
\lim _{x \to \pm\infty }\phi (t,x) = \pm1\:(=\mp 1), \quad
\lim _{x \to \pm\infty }\xi (t,x) = 0 \quad
\forall\ t\in{\bf R},
\nonumber
\end{eqnarray}
i.e., when $\phi(t,x)$ has the form of a transition layer between two different vacua
such that $P\ne 0$ (in contrast to a nontopological soliton, when the solution tends to
the same vacuum value as $x \to \pm\infty$, i.e., when $\phi(t,x)$ has the form of a
splash over the true vacuum such that $P=0$).

In these terms, solution (\ref{wall}) is a topological soliton with zero charge $Q$.

{\bf Definition 2.} A topological soliton is said to additionally bear a $U(1)$-charge
if $Q \neq 0$. Such a soliton is reffered to as a {\it topological Q-ball} (in contrast to
a nontopological Q-ball with $P=0$).

The conditions for the existence of nontopological Q-balls in the Lee-Friedberg-Sirlin
model~\cite{FLS} and their stability are discussed, along with \cite{FLS}, in~\cite{1} Chapter~10,
and in \cite{2}. In particular, it is shown that stable nontopological spherically symmetric
and one-dimensional Q-balls for such a model exist for large values of the charges,
$Q>Q_{c}$, where $Q_c$ is a critical charge depending on the parameters of the Lagrangian.
However, the explicit form of such Q-balls has not yet been found.

In \cite{2}, for system (\ref{eulereq1}), (\ref{eulereq2}), the following precise solution
of the topological Q-ball type was found:
\begin{equation}
\phi_0(t,x)\equiv\phi_0(x)=\tanh(x),
\label{phisol}
\end{equation}
\begin{equation}
\xi_0(t,x)=\displaystyle\sqrt{\frac{1}{\kappa^2}-\frac{1}{2}}
\frac{\exp(it)}{\cosh(x)},\ \
0<\kappa^2<2, \quad t, x \in {\bf R}.
\label{xisol}
\end{equation}
For this configuration, from (\ref{energy}) -- (\ref{tcharge}), we obtain
\begin{equation}
E_0=\frac{4}{3}\left(\frac{4}{\kappa^2}-1\right),
\label{Esol}
\end{equation}
\begin{equation}
Q_0=2\left(\frac{2}{\kappa^2}-1\right),
\label{Qsol}
\end{equation}
\begin{equation}
P_0=1 .
\label{Psol}
\end{equation}
When $\kappa\to\sqrt{2}$ the solution (\ref{phisol}), (\ref{xisol}) tends to (\ref{wall}).

{\bf Remark 1.} It follows from the above discussions, that the functions
$\tilde{\phi}_0=-\phi_0$ and $\tilde{\xi}_0=\xi_0\exp(i\alpha)$ are also
solutions to system (\ref{eulereq1}), (\ref{eulereq2}) with the same $E_0$, $Q_0$,
and $\tilde{P}_0=-1$.

{\bf Remark 2.} Taking into account the invariance of Eqs. (\ref{eulereq1}), (\ref{eulereq2})
with respect to the Lorentz transformations of the independent variables $t$ and $x$,
we find that solution (\ref{wall}) generates for Eq. (\ref{eulereq2}) (for $\xi\equiv 0$)
the traveling wave front $\phi_{w}(t,x)=\tanh[(\sqrt{2}/\kappa)(x\pm vt)/\sqrt{1-v^2}]$
if an initial speed $v$: $0<v<1$, is given; solution (\ref{phisol}), (\ref{xisol}) generates
for system (\ref{eulereq1}), (\ref{eulereq2}) the traveling wave of the form
$\phi_Q(t,x)=\tanh\left[(x\pm vt)/\sqrt{1-v^2}\right]$,
$\xi_Q(t,x)=\displaystyle\sqrt{\frac{1}{\kappa^2}-\frac{1}{2}}\:\displaystyle\frac{\exp[i(t\pm vx)/\sqrt{1-v^2}]}{\cosh[(x\pm vt)/\sqrt{1-v^2}]}$
with oscillations in the $\xi$-component.

In conclusion of this section, we briefly discuss physical interpretation (in addition to that
in~\cite{2}) of solution (\ref{phisol}), (\ref{xisol}). This solution possesses properties
similar to those of possible Q-balls~\cite{Col} in the (3+1)-dimensional Minkowski space,
which are assumed to have something to do with the problem of the baryon asymmetry
of the Universe. One of the mechanisms that can explain the baryon asymmetry has been
suggested in~\cite{AfDi}. In accordance with this mechanism, at the late inflation stages,
a condensate is formed that can evolve into Q-balls that bear the same baryon charge
but are more advantageous from the energy standpoint. Note that these Q-balls may
continue to exist until now and contribute to the dark matter. Moreover, it has been
noted in~\cite{KuAl} that such ''relict'' Q-balls may occur crucial factors when studying
stability of neutron stars.

The precise solution of the Q-ball type in the (1+1)-dimensional Minkowski space,
which was found in~\cite{2} and is studied in this paper, is important not only as
an approximation of (3+1)-dimensional Q-balls; it is of great interest in connection
with studying domain walls and processes on them.

\section{Dynamic stability of the solution and spectral problem statement}
\setcounter{equation}{0}

From the standpoint of physical applications, the problem of the dynamic stability
of solution (\ref{phisol}), (\ref{xisol}) is of great importance. A solution is considered
to be absolutely stable if the energy functional takes its absolute minimum on this
solution. For system (\ref{eulereq1}), (\ref{eulereq2}) true vacua (\ref{truevacs})
are solutions of this kind. However, these solutions possess zero $Q$- and $P$-charges.

{\bf Definition 3.} A solution $\xi(t,x)$, $\phi(t,x)$ to system (\ref{eulereq1}), (\ref{eulereq2})
is said to be {\it absolutely stable} in a sector $\{P, Q\}$ if it has the least energy among
all solutions with fixed values of charges (\ref{charge}) and (\ref{tcharge}).

Solution (\ref{wall}) is absolutely stable in the sector $\{1,0\}$, since it is the only
stationary solution of system (\ref{eulereq1}), (\ref{eulereq2}) in this sector.

The question of whether solution (\ref{phisol}), (\ref{xisol}) is absolutely stable in the sector
$\{P_0, Q_0\}$, where $Q_0$ and $P_0$ are defined by (\ref{Qsol}) and (\ref{Psol}), respectively,
is not easy to answer because of the dependence of component (\ref{xisol}) on time.

A suggestive consideration regarding the stability of solution (\ref{phisol}), (\ref{xisol})
might be as follows: the parameter $\kappa^2$ affects only the amplitude in (\ref{xisol});
for $\kappa^2=2$, solution (\ref{phisol}), (\ref{xisol}) turns to the absolutely stable solution
(\ref{wall}); and, for $0<\kappa^2<2$, there appears a nonzero $U(1)$-charge, which
usually only stabilizes the solution (see \cite{1}, Chapter 10).

The main part of this paper is devoted to studying the dynamic stability of
solution (\ref{phisol}), (\ref{xisol}) with respect to small perturbations
(Lyapunov's stability in the framework of linear theory). However, first,
we present some physical considerations regarding the stability of
configuration (\ref{phisol}), (\ref{xisol}) from the standpoint of its possible
disintegration into charged nonlocalized formations.

1. The problem consists in searching for possible solutions in the sector
$\{P_0, Q_0\}=\{1,Q_0\}$ that are close to solution (\ref{wall}), which is
absolutely stable in the sector $\{1,0\}$. The solutions are sought in the form
of small nonlocalized perturbations of the field $\xi$ determining the charge
$Q_0$. Such solutions, if they exist, may occur equivalent to or more
advantageous than solution (\ref{phisol}), (\ref{xisol}) from the energy
considerations. An approximate solution to system (\ref{eulereq1}), (\ref{eulereq2})
close to (\ref{wall}) is sought in the form $\{\delta\xi(t,x),\phi_w(x)\}$, where $|\delta\xi|\ll 1$
and
\begin{equation}
\phi_w=\tanh\left(\sqrt{2}\:x/\kappa\right),
\label{mute}
\end{equation}
i.e., only the $\xi$-component in (\ref{wall}) is perturbed. From (\ref{eulereq1}),
we obtain the following equation in $\delta\xi$:
\begin{equation}
\frac{\partial^2\delta\xi}{\partial t^2}-\frac{\partial^2\delta\xi}{\partial x^2}
+2\left[1-\frac{1}{\cosh^2\left(\sqrt{2}\:x/\kappa\right)}\right]\delta\xi=0.
\label{eq1}
\end{equation}
To separate variables in (\ref{eq1}), the solution is sought in the form
\begin{equation}
\delta\xi(t,x)=e^{i\omega t}f(x),
\label{harm}
\end{equation}
where $\omega$ is a separation parameter and $f(x)$ satisfies the ODE
\begin{equation}
\frac{d^2 f}{dx^2}+\left[\omega^2-2+\frac{2}{\cosh^2\left(\sqrt{2}\:x/\kappa\right)}\right]f=0, \quad x\in {\bf R}.
\label{eq3}
\end{equation}
For reasons that will be clear from the following discussions, we assume that
the solution is defined in an arbitrarily large but still finite interval $(-L,L)$.
From (\ref{energy}) and (\ref{charge}), we obtain the following expressions for
the charge and energy of solution (\ref{harm}):
\begin{equation}
Q=2\omega\int\limits_{-L}^{L}|f(x)|^2\:dx,
\label{norm}
\end{equation}
\begin{equation}
\delta E=\int\limits_{-L}^{L}\left[\omega^2 |f|^2+\left|\frac{df}{dx}\right|^2
+2\tanh^2\left(\sqrt{2}\:x/\kappa\right)|f|^2\right]dx.
\label{eq4}
\end{equation}

In the interval $0\le\omega^2<2$, Eq. (\ref{eq3}) may have only a descrete spectrum.
If $\omega^2$ is an eigenvalue from the discrete spectrum, the corresponding
eigenfunction $f(x,\omega^2)$ belongs to $L_2(-\infty,\infty)$ and has the following
asymptotics for large $|x|$:
\begin{equation}
f(x,\omega^2)\sim
A\exp\left(-\sqrt{2-\omega^2}\:|x|\right),
\label{asym1}
\end{equation}
where $A$ is a normalization constant. Since the perturbation must be
normalized by the given charge $Q_0$, the localized eigenfunction with
asymptotics (\ref{asym1}) does not generally meet the requirement of the
perturbation smallness for large $Q_0$.

The values $\omega^2>2$ belong to the continuous spectrum. If $f(x,\omega^2)$
is an eigenfunction corresponding to an eigenvalue belonging to the
continuous spectrum, it has the following asymptotics for large $|x|$:
\begin{equation}
f(x,\omega^2)\sim
A\:\cos\left(\sqrt{\omega^2-2}\:|x|+\delta\right),
\label{asym2}
\end{equation}
where the phase $\delta$ is uniquely determined by the boundary condition
at $x=0$, either $f(0)=0$ or $f^{\prime}(0)=0$, and the amplitude $A$ is the
normalization constant. If normalization (\ref{norm}) is used, the amplitude $A$
may be selected as small as desired through an appropriate choice of $L$.
In addition, $f^{\prime}(x,\omega^2)\sim 0$ for $\omega^2\sim 2$, so that the least
value for (\ref{eq4}) is obtained when $\omega^2\to 2$ in (\ref{asym2}).
Since, in the framework of the approximation considered, integrals
(\ref{energy}) and (\ref{charge}) with the infinite limits diverge on solutions
(\ref{mute}), (\ref{harm}), and (\ref{asym2}) for $\omega^2\ge 2$, the function $f(x,\omega^2)$
is set equal to zero outside some arbitrarily large (but finite) interval
$-L<x<L$. Note that this expedient is often used when solving physical problems
(see, for example, \cite{LaLi}, pp. 170-174; \cite{1}, Chapter 10.)

Let us multiply (\ref{eq3}) by $f^{*}(x)$ and integrate the equation obtained
in the interval $-L\le x\le L$. In so doing, the term $\displaystyle\frac{d^2f}{dx^2}f^{*}$
is integrated by parts with regard to the conditions $f(\pm L)=0$.
As a result, we obtain
\begin{eqnarray}
\int\limits_{-L}^{L}\left[-\omega^2 |f|^2+\left|\frac{df}{dx}\right|^2
+2\tanh^2\left(\sqrt{2}\:x/\kappa\right)|f|^2\right]dx=0.
\nonumber
\end{eqnarray}
From this and (\ref{eq4}), we have
\begin{eqnarray}
\delta E=2\omega^2\int\limits_{-L}^{L}|f|^2dx.
\nonumber
\end{eqnarray}
Hence, with regard to (\ref{norm}), it follows that
\begin{equation}
\delta E=\omega Q\ge
\sqrt{2}\:Q
\label{eq7}
\end{equation}
for $\omega^2\ge 2$. Finally, we find that the energy of the configuration composed
of the domain wall (\ref{mute}) with charged perturbations (\ref{harm}) over it is
the sum of $E_w$ and $\delta E$ given by (\ref{Ew}) and (\ref{eq7}), respectively; i.e.,
\begin{equation}
E=\omega Q+4\sqrt{2}/(3\kappa)\ge\sqrt{2}Q+4\sqrt{2}/(3\kappa), \quad \omega\ge\sqrt{2}.
\label{eq10}
\end{equation}
Note that the energy $E_0$ and charge $Q_0$ of the precise solution
(\ref{phisol}), (\ref{xisol}) are uniquely specified by the parameters of Lagrangian (\ref{lagr}).
At the same time, in the linear approximation, configuration (\ref{mute}), (\ref{harm})
may have arbitrary $E$ and $Q$ satisfying (\ref{eq10}).

Inequality (\ref{eq10}) can be derived in a different way. Since the domain wall
(\ref{mute}) is localized on the interval of length $\sim\kappa/\sqrt{2}$, to compute
the energy of nonlocalized perturbations of the field $\xi$, we substitute
$\phi^2=1$ and $\xi=\delta\xi(t,x)$ into Eq. (\ref{eulereq1}), which implies the
perturbation of vacua (\ref{truevacs}) along the component $\xi$ (here, we
rely on the physically reasonable assumption that the field $\xi$ is concentrated
basically outside the domain where the wall is localized). For the function
$\delta\xi$, we obtain the equation (cf. with (\ref{eq1}))
\begin{equation}
\frac{\partial^2\delta\xi}{\partial t^2}-\frac{\partial^2\delta\xi}{\partial x^2}
+2\delta\xi=0.
\label{lineq}
\end{equation}
In order to avoid diverging integrals in the computation of the energy and charge,
we again consider the perturbations $\delta\xi$ on a large, but finite, interval
$-L<x<L$. Among solutions of Eq. (\ref{lineq}) corresponding to a given charge
$Q$, the normalized solution
\begin{equation}
\delta\xi=\sqrt{Q/(4\sqrt{2}L)}\exp{(i\sqrt{2}\:t)},
\label{xiosc}
\end{equation}
has the least energy (note that this solution does not depend on $x$).
Here, the normalization multiplier is determined from the condition that
the U(1)-charge for the field $\delta\xi$ is equal to $Q$, i.e.,
$$
-i\displaystyle\int\limits_{-L}^{L}\left(\delta\xi^*\frac{\partial\delta\xi}{\partial t}-
\delta\xi\frac{\partial\delta\xi^*}{\partial t}\right)dx=Q.
$$
The energy of solution (\ref{xiosc}) is equal to $\sqrt{2}Q$. Adding together
this quantity and the wall energy (\ref{Ew}), we obtain the expression coinciding
with the right-hand side of (\ref{eq10}).

For the precise solution (\ref{phisol}), (\ref{xisol}), we obtain from (\ref{Esol})
and (\ref{Qsol}) the following relationship between the charge and energy:
\begin{equation}
E_0=4(Q_0 + 1)/3.
\label{eq11}
\end{equation}
It is easy to see from (\ref{eq10}) and (\ref{eq11}) that, for any admissible
values of the parameter $\kappa$, the energy of the precise solution is
less than the minimal possible energy of the nonlocalized configuration
(\ref{mute}), (\ref{harm}) with the same charge. This brings us to the conclusion
that solution (\ref{phisol}), (\ref{xisol}) is stable in terms of the disintegration
into a nonlocalized configuration of the type ''wall + plane waves''.

2. The basic objective of this work is to study the dynamic stability of
solution (\ref{phisol}), (\ref{xisol}) in the framework of linear perturbation
theory. We use the approach similar to that employed in studying
the stability of localized solutions of some nonlinear wave equations
in field theory \cite{3} -- \cite{5}. In particular, it was used in \cite{5} for
studying the stability of a precise nontopological soliton carrying
a U(1)-charge for a complex wave equation with fifth-degree nonlinearity.

Let us set $\phi=\phi_0+\delta\phi$ and $\xi=\xi_0+\delta\xi$, where
$\delta\phi$ and $\delta\xi$ are small deviations from (\ref{phisol}),
(\ref{xisol}); note that $\delta\phi(t,x)$ is a real-valued function and
$\delta\xi(t,x)$ is a complex-valued function. System (\ref{eulereq1}),
(\ref{eulereq2}) reduces to the following linearized system of the
differential equations in the deviations:
\begin{equation}
\frac{\partial^2\delta\xi}{\partial t^2}-
\frac{\partial^2\delta\xi}{\partial
x^2}+2\phi_0^2\delta\xi+4\phi_0\xi_0\delta\phi=0,
\label{lineqxi}
\end{equation}
\begin{equation}
\frac{\partial^2\delta\phi}{\partial t^2}-
\frac{\partial^2\delta\phi}{\partial
x^2}+4\xi_0\xi^*_0\delta\phi+4\phi_0(\xi_0\delta
\xi^*+\xi_0^*\delta\xi)+\frac{4}{\kappa^2}(3\phi_0^2-
1)\delta\phi=0.
\label{lineqphi}
\end{equation}
Similarly \cite{3} -- \cite{5}, the solution to Eqs. (\ref{lineqxi}), (\ref{lineqphi})
is sought in the form
\begin{equation}
\delta\xi=[\eta(x)\exp(-i\lambda t)+\chi^*(x)\exp(i\lambda^*t)]\exp(it),
\label{deltaxi}
\end{equation}
\begin{equation}
\delta\phi=V(x)\exp(-i\lambda t)+V^*(x)\exp(i\lambda^*t).
\label{deltaphi}
\end{equation}
Such a representation of the solution makes it possible to separate
variables in (\ref{lineqxi}), (\ref{lineqphi}) and to obtain ODEs for
the amplitudes of the perturbations $\eta, \chi$, and $V$ not depending
on $t$. Indeed, substituting (\ref{phisol}), (\ref{xisol}), (\ref{deltaxi}),
and (\ref{deltaphi}) into (\ref{lineqxi}) and (\ref{lineqphi}), we obtain
the following system of ODEs for $\eta$, $\chi$, and $V$ depending
on the parameter $\lambda$:
\begin{equation}
\eta^{\prime\prime}=\left(1+2\lambda-\lambda^2-
\frac{2}{\cosh^2{x}}\right)\eta
+4\:\sqrt{\frac{1}{\kappa^2}-\frac{1}{2}}\:\frac{\tanh{x}}{\cosh{x}}\:V,
\label{system1}
\end{equation}
\begin{equation}
\chi^{\prime\prime}=\left(1-2\lambda-\lambda^2-
\frac{2}{\cosh^2{x}}\right)\chi
+4\:\sqrt{\frac{1}{\kappa^2}-\frac{1}{2}}\:\frac{\tanh{x}}{\cosh{x}}\:V,
\label{system2}
\end{equation}
\begin{equation}
V^{\prime\prime}=\left(\frac{8}{\kappa^2}-\lambda^2-
\frac{8+2\kappa^2}{\kappa^2\cosh^2{x}}\right)V
+4\:\sqrt{\frac{1}{\kappa^2}-
\frac{1}{2}}\:\frac{\tanh{x}}{\cosh{x}}\:(\eta+\chi),\quad x\in\mbox{\bf R}\ .
\label{system3}
\end{equation}
Solutions to system (\ref{system1})~-- (\ref{system3}) are sought in the class
of square integrable functions defined on the entire real axis satisfying
the conditions
\begin{eqnarray}
\lim _{x\to\pm\infty}\eta (x) =
\lim _{x\to\pm\infty}\chi (x) =
\lim _{x\to\pm\infty}V(x) =
\nonumber
\end{eqnarray}
\begin{equation}
=\lim _{x\to\pm\infty}\eta^{\prime}(x) =
\lim _{x\to\pm\infty}\chi^{\prime}(x) =
\lim _{x\to\pm\infty}V^{\prime}(x) = 0.
\label{bconds}
\end{equation}
It is required to find the values of $\lambda$ (eigenvalues) for which the singular
boundary value problem (\ref{system1})~-- (\ref{bconds}) has nontrivial solutions
(eigenfunctions). By virtue of (\ref{deltaxi}), (\ref{deltaphi}), for any complex
eigenvalue $\lambda$ with a nonzero imaginary part, the perturbations grow
exponentially in time. Hence, the dynamic stability of solution (\ref{phisol}),
(\ref{xisol}) with respect to small perturbations of form (\ref{deltaxi}), (\ref{deltaphi})
requires that the discrete spectrum of problem (\ref{system1})~-- (\ref{bconds})
be real.

In addition, if the ODE system (\ref{system1}) -- (\ref{system3}) has a continuous
spectrum, it also must lie on the real axis of the complex plane $\lambda$
to make solution (\ref{phisol}), (\ref{xisol}) stable.

\section{Analytic properties of the spectral problem}
\setcounter{equation}{0}

Taking into account that the ODE system (\ref{system1})~-- (\ref{system3})
is not changed upon the replacement of ($x$, $\eta$, $\chi$, $V$) by ($-x$, $\eta$, $\chi$, $-V$)
or ($-x$, $-\eta$, $-\chi$, $V$), we can consider the singular boundary value problem
on the semiaxis and write it in the following final form:
\begin{equation}
\eta^{\prime\prime}=\left(1+2\lambda-\lambda^2-
\frac{2}{\cosh^2{x}}\right)\eta
+4\:\sqrt{\frac{1}{\kappa^2}-\frac{1}{2}}\:\frac{\tanh{x}}{\cosh{x}}\:V,
\label{equ1}
\end{equation}
\begin{equation}
\chi^{\prime\prime}=\left(1-2\lambda-\lambda^2-
\frac{2}{\cosh^2{x}}\right)\chi
+4\:\sqrt{\frac{1}{\kappa^2}-\frac{1}{2}}\:\frac{\tanh{x}}{\cosh{x}}\:V,
\label{equ2}
\end{equation}
\begin{equation}
V^{\prime\prime}=\left(\frac{8}{\kappa^2}-\lambda^2-
\frac{8+2\kappa^2}{\kappa^2\cosh^2{x}}\right)V
+4\:\sqrt{\frac{1}{\kappa^2}-\frac{1}{2}}\:\frac{\tanh{x}}{\cosh{x}}\:(\eta+\chi),
\label{equ3}
\end{equation}
\begin{eqnarray}
0\le x<\infty, \quad \kappa^2\le 2,
\nonumber
\end{eqnarray}
\begin{equation}
\eta(0)=\chi(0)=V^{\prime}(0)=0 \quad \mbox{or} \quad
\eta^{\prime}(0)=\chi^{\prime}(0)=V(0)=0,
\label{cond0}
\end{equation}
\begin{equation}
\lim_{x\to\infty}\eta(x)=\lim_{x\to\infty}\eta^{\prime}(x)=0,
\label{condinf1}
\end{equation}
\begin{equation}
\lim_{x\to\infty}\chi(x)=\lim_{x\to\infty}\chi^{\prime}(x)=0,
\label{condinf2}
\end{equation}
\begin{equation}
\lim_{x\to\infty}V(x)=\lim_{x\to\infty}V^{\prime}(x)=0.
\label{condinf3}
\end{equation}
This problem contains terms depending nonlinearly on the spectral
parameter $\lambda$ and, generally, may have complex eigenvalues.
We seek for nontrivial solutions to this problem in the class of
complex-valued functions satisfying the condition
\begin{equation}
\int\limits_0^{\infty}\left[\eta^{*}(x)\eta(x)+\chi^{*}(x)\chi(x)+V^{*}(x)V(x)\right]dx<\infty.
\label{condint}
\end{equation}

Note that the singular boundary value problem (\ref{equ1}) -- (\ref{condinf3})
is correctly defined in terms of the number of the boundary conditions
for large $x$ if and only if the singular Cauchy problem (\ref{equ1}) -- (\ref{equ3}),
(\ref{condinf1}) -- (\ref{condinf3}) admits a three-parameter family of solutions
at infinity. Since system (\ref{equ1}) -- (\ref{equ3}) is asymptotically equivalent
to a system with constant coefficients, each of the decoupled second-order
ODE for large $x$ must have a one-parameter family of solutions vanishing
at infinity. This results in the following requirements on the location of the
eigenvalues $\lambda$ in the complex plane:
\begin{equation}
1+2\lambda-\lambda^2\notin(-\infty, 0], \quad
1-2\lambda-\lambda^2\notin(-\infty, 0], \quad
8/\kappa^2-\lambda^2\notin(-\infty, 0],
\label{condlambda}
\end{equation}
i.e., they do not lie on the nonpositive real semiaxis. Then, for sufficiently
large $x$, $x\gg 1$, the three-dimensional linear subspace generated
in the phase space {\bf C}$^6$ of system (\ref{equ1}) -- (\ref{equ3}) by the
values of the solutions of the singular Cauchy problem (\ref{equ1}) -- (\ref{equ3}),
(\ref{condinf1}) -- (\ref{condinf3}) is given in the form
\begin{equation}
\eta^{\prime}(x)\approx-\sqrt{1+2\lambda-\lambda^2}\:\eta(x),
\label{approx1}
\end{equation}
\begin{equation}
\chi^{\prime}(x)\approx-\sqrt{1-2\lambda-\lambda^2}\:\chi(x),
\label{approx2}
\end{equation}
\begin{equation}
V^{\prime}(x)\approx-\sqrt{8/\kappa^2-\lambda^2}\:V(x),
\label{approx3}
\end{equation}
where the roots are assumed to have positive real parts (see \cite{6} for detail).

Hence, the following assertions are valid:

(1) real eigenvalues of problem (\ref{equ1}) -- (\ref{condint}), if exist,
satisfy (by virtue of (\ref{condlambda})) the inequalities
\begin{equation}
-\sqrt{2}+1<\lambda<\sqrt{2}-1;
\label{estim1}
\end{equation}

(2) the geometric multiplicity of any eigenvalue of problem (\ref{equ1}) -- (\ref{condint})
cannot be greater than three; i.e., not more than three linearly independent
eigenfunctions may correspond to one eigenvalue;

(3) system (\ref{equ1}) -- (\ref{equ3}) has a continuous spectrum lying
on the real axis of the complex plane $\lambda$ in the intervals
$(-\infty, -\sqrt{2}+1)$ and $(\sqrt{2}-1, \infty)$ (conditions (\ref{approx1}),
(\ref{approx2}), and (\ref{approx3}) turn to conditions of the radiation
type for $\lambda\in (-\infty, -\sqrt{2}+1)$ and 
$\lambda\in (1+\sqrt{2}, \infty)$, for $\lambda\in (-\infty, -\sqrt{2}-1)$ and
$\lambda\in (\sqrt{2}-1, \infty)$, and for $\lambda^2>8/\kappa^2\ge 4$, respectively);
these real intervals of the continuous spectrum do not affect stability
of solution (\ref{phisol}), (\ref{xisol}) with respect to small perturbations
of form (\ref{deltaxi}), (\ref{deltaphi}).

Note also that, if $\lambda$ is a purely imaginary eigenvalue of problem
(\ref{equ1}) -- (\ref{condint}), then $\eta=\chi^{*}$, and $V$ is a real-valued
function.

These assertions can be supplemented by the following facts and estimates.

1. {\bf Case of $\kappa^2$=2.} For $\kappa^2=2$, system (\ref{equ1}) -- (\ref{equ3})
is decoupled into the three second-order ODEs:
\begin{equation}
\eta^{\prime\prime}+\left[2/\cosh^2{x}-(1+2\lambda-\lambda^2)\right]\eta=0,
\label{equ1a}
\end{equation}
\begin{equation}
\chi^{\prime\prime}+\left[2/\cosh^2{x}-(1-2\lambda-\lambda^2)\right]\chi=0,
\label{equ2a}
\end{equation}
\begin{equation}
V^{\prime\prime}+\left[6/\cosh^2{x}-(4-\lambda^2)\right]V=0,
\label{equ3a}
\end{equation}
each of which can be reduced to a hypergeometric equation. Indeed,
consider the equation
\begin{equation}
\frac{d^2\psi}{dx^2}+\left[\frac{s(s+1)}{\cosh^2{x}}-\varepsilon^2\right]\psi=0, \quad
-\infty<x<\infty,
\label{hyper1}
\end{equation}
and introduce the change of variable $y=\tanh x$. For the function $\psi(y)$,
we obtain the equation
\begin{eqnarray}
\frac{d}{dy}\left[\left(1-y^2\right)\frac{d\psi}{dy}\right]+
\left[s(s+1)-\frac{\varepsilon^2}{1-y^2}\right]\psi=0, \quad
-1<y<1,
\nonumber
\end{eqnarray}
which reduces to a hypergeometric equation by means of the substitution
$\psi(y)=\left(1-y^2\right)^{\varepsilon/2}w(y)$ and the change of variable
$(1-y)/2=z$. Assuming that $w=w(z)$, we obtain
\begin{eqnarray}
\begin{array}{c}
z(1-z)w^{\prime\prime}+
[(\varepsilon+1)-((\varepsilon-s)+(\varepsilon+s+1)+1)z]w^{\prime}-
(\varepsilon-s)(\varepsilon+s+1)w=0,\\
0<z<1.
\end{array}
\nonumber
\end{eqnarray}
This equation has the two independent integrals \cite{Lebedev}:
\begin{eqnarray}
w_1(z)=F(\varepsilon-s,\ \varepsilon+s+1,\ \varepsilon+1,\ z), \quad
w_2(z)=z^{-\varepsilon}F(s+1,\ -s,\ 1-\varepsilon,\ z),
\nonumber
\end{eqnarray}
where $F$ is the hypergeometric function
\begin{eqnarray}
F(\alpha, \beta, \gamma, z)=
\sum_{k=0}^{\infty}\frac{(\alpha)_k(\beta)_k}{(\gamma)_k k!}z^k,
\quad
(\sigma)_k=\frac{\Gamma(\sigma+k)}{\Gamma(\sigma)}.
\nonumber
\end{eqnarray}
Hence, returning to (\ref{hyper1}), we obtain
\begin{eqnarray}
\psi_1(x)=\frac{1}{\cosh^{\varepsilon}x}
\ F\left(\varepsilon-s,\ \varepsilon+s+1,\ \varepsilon+1,\ (1-\tanh x)/2\right),
\nonumber
\end{eqnarray}
\begin{eqnarray}
\psi_2(x)=2^{\varepsilon}\exp{(\varepsilon}x)
\ F\left(s+1,\ -s,\ 1-\varepsilon,\ (1-\tanh x)/2\right).
\nonumber
\end{eqnarray}
Then, only $\psi_1$ belongs to $L_2(-\infty,\infty)$ under the condition
that $\varepsilon>0$ and $\varepsilon-s=-n$, where $n=0,\ 1,\ 2,\ \dots$,
when $F$ becomes a polynomial of the $n$-th degree in $\tanh x$ and,
hence, has finite limits as $x\to\pm\infty$.

For example, for Eq. (\ref{equ1a}), we have $s=1$, and
$\varepsilon=\sqrt{1+2\lambda-\lambda^2}=1-n>0$. Hence, it follows that
$n=0$, $\lambda_1=0$, $\lambda_2=2$, and
$\psi_1(x,\lambda_1)=\psi_1(x,\lambda_2)\equiv\eta(x)=1/\cosh x$.
Equations (\ref{equ2a}) and (\ref{equ3a}) are treated in a similar way.
Hence, in the case of $\kappa^2=2$, all points of the discrete spectrum
of problem (\ref{equ1}) -- (\ref{condint}) and all eigenfunctions corresponding
to them can be found (all eigenfunctions are accurate up to the normalizing
multipliers):
\begin{equation}
\lambda_1=0: \quad \eta\equiv 0, \quad \chi\equiv 0, \quad V=1/\cosh^2 x;
\label{eig1}
\end{equation}
\begin{equation}
\lambda_1=0\ \mbox{and}\ \lambda_2=2: \quad \eta=1/\cosh x, \quad \chi\equiv 0, \quad V\equiv 0;
\label{eig2}
\end{equation}
\begin{equation}
\lambda_1=0\ \mbox{and}\ \lambda_3=-2: \quad \eta\equiv 0, \quad \chi=1/\cosh x, \quad V\equiv 0;
\label{eig3}
\end{equation}
\begin{eqnarray}
\lambda_{4,5}=\pm\sqrt{3}: \quad \eta\equiv 0, \quad \chi\equiv 0, \quad V=\tanh x/\cosh x.
\nonumber
\end{eqnarray}
Thus, for $\kappa^2=2$, the eigenvalue $\lambda=\lambda_1=0$ of problem
(\ref{equ1}) -- (\ref{condint}) has the maximal possible geometric multiplicity
equal to three, whereas the multiplicity of the eigenvalues $\lambda_2$,
$\lambda_3$, $\lambda_4$, and $\lambda_5$ is equal to one.

2. {\bf Case of 0<$\kappa^2$<2.}
It is easy to check that, for any $\kappa$ satisfying the inequalities $0<\kappa^2<2$,
$\lambda=\lambda_1=0$ remains an eigenvalue of problem (\ref{equ1}) -- (\ref{condint}),
and there are, at least, two eigenfunctions corresponding to it; namely,
\begin{equation}
\lambda_1=0: \quad \eta=-\chi=1/\cosh x, \quad V=0;
\label{eig5}
\end{equation}
\begin{equation}
\lambda_1=0: \quad \eta=\chi=-\sqrt{\frac{1}{\kappa^2}-\frac{1}{2}}\frac{\tanh{x}}{\cosh{x}}, \quad V=\frac{1}{\cosh^2{x}}.
\label{eig6}
\end{equation}

Note further that, if $(\eta(x), \chi(x), V(x), \lambda)$ is a solution to problem
(\ref{equ1}) -- (\ref{condint}), then $(\eta^*(x), \chi^*(x), V^*(x), \lambda^*)$,
$(\chi(x), \eta(x), V(x), -\lambda)$ and $(\chi^*(x), \eta^*(x), V^*(x), -\lambda^*)$
are also its solutions, so that it is sufficient to consider, for example,
the following range of $\lambda$:
\begin{equation}
\mbox{Re}\:\lambda\ge 0, \quad \mbox{Im}\:\lambda\ge 0.
\label{sector}
\end{equation}
Like in \cite{3} -- \cite{5}, we write system (\ref{equ1}) -- (\ref{equ3}) in the form
\begin{equation}
\lambda^2I\Psi-2\lambda D\Psi-H\Psi=0, \quad 0\le x<\infty,
\label{sysmatrix}
\end{equation}
where $\Psi=(\eta, \chi, V)^T$ ($T$ denotes the transposition) and $I$, $H$, and $D$
are operators defined in the space of three-component complex-valued twice
differentiable functions $\Psi$ bounded on the semiaxis and satisfying conditions
(\ref{cond0}) -- (\ref{condint}), $\Psi$: $[0,\infty)\longrightarrow{\bf C}^3$, with the scalar
product defined as
$$(\Psi,\Psi)=\displaystyle\int\limits_0^{\infty}\Psi^{*}(x)\Psi(x)\:dx.$$
Here, $I$ is the identity operator and $D$ and $H$ have the form
\begin{equation}
D=
\left\|
\begin{array}{ccc}
1&0&0\\
0&-1&0\\
0&0&0\\
\end{array}
\right\|,
\qquad
H=
\left\|
\begin{array}{ccc}
H_{11}&0&H_{13}\\
0&H_{22}&H_{23}\\
H_{31}&H_{32}&H_{33}\\
\end{array}
\right\|,
\label{opers}
\end{equation}
where
\begin{equation}
\begin{array}{c}
H_{11}=H_{22}=\displaystyle-\frac{d^2}{dx^2}-
\frac{2}{\cosh^2{x}}+1,\\
\\
H_{33}=\displaystyle-\frac{d^2}{dx^2}-
\left(\frac{8}{\kappa^2}+2\right)
\frac{1}{\cosh^2{x}}+\frac{8}{\kappa^2},\\
\\
H_{13}=H_{31}=H_{23}=H_{32}=4\:\displaystyle\sqrt{\frac{1}{\kappa^2}-\frac{1}{2}}\:\frac{\tanh{x}}{\cosh{x}}.
\end{array}
\label{H}
\end{equation}
Taking the scalar product of (\ref{sysmatrix}) with $\Psi$ and assuming that $\Psi$
is normalized such that $(\Psi,\Psi)=1$), we obtain
\begin{equation}
\lambda^2-2\lambda\left(\Psi,D\Psi\right)-
\left(\Psi,H\Psi\right)=0.
\label{squeq}
\end{equation}
$H$ and $D$ are Hermitian operators. The operator  $D$ is obviously Hermitian;
$H$ is Hermitian because the operator $-d^2/dx^2$ is Hermitian. The latter follows
from the fact that the substitutions $\Psi^{*}\displaystyle\left.\frac{d\Psi}{dx}\right|_0^{\infty}$
and $\displaystyle\left.\frac{d\Psi^{*}}{dx}\Psi\right|_0^{\infty}$ turn to zero on the
functions satisfying conditions (\ref{cond0}) -- (\ref{condinf3}). Then, $(\Psi,D\Psi)$ and
$(\Psi,H\Psi)$ are real numbers. From (\ref{squeq}), it follows that
\begin{equation}
\lambda=\displaystyle\left(\Psi,D\Psi\right)
\pm\sqrt{\displaystyle\left(\Psi,D\Psi\right)^2
+\left(\Psi,H\Psi\right)}.
\label{roots}
\end{equation}
From this equation, we can obtain estimates for an eigenvalue with a nonzero
imaginary part. Indeed, in this case, it follows from (\ref{roots}) that
\begin{equation}
\mbox{Re}\:\lambda=\left(\Psi,D\Psi\right),
\quad
\mbox{Im}\:\lambda=\sqrt{-
\displaystyle\left(\Psi,D\Psi\right)^2
-\left(\Psi,H\Psi\right)},
\quad
|\lambda|^2=-\left(\Psi,H\Psi\right) .
\label{ReImAbsLambda}
\end{equation}
We have the following estimates for the real and imaginary parts of $\lambda$:
\begin{eqnarray}
|\mbox{Re}\:\lambda|=|(\Psi, D\Psi)|=
\left|\displaystyle\int\limits^{\infty}_{0}dx\left(\eta^*\eta-\chi^*\chi\right)\right|\le\nonumber\\
\le\displaystyle\int\limits^{\infty}_{0}dx\left|\eta^*\eta-\chi^*\chi\right|\le
\displaystyle\int\limits^{\infty}_{0}dx\left(\eta^*\eta+\chi^*\chi\right)
\le 1;
\label{ReLambda}
\end{eqnarray}
\begin{eqnarray}
|\mbox{Im}\:\lambda|=\sqrt{-\left(\Psi,D\Psi\right)^2
-\left(\Psi,H\Psi\right)}\le\sqrt{-(\Psi, H\Psi)}\le\sqrt{-\mu_{min}}\ ,
\label{ImLambda}
\end{eqnarray}
where $\mu_{min}$ is the least eigenvalue of the operator $H$. Let us obtain
an estimate for $\mu_{min}$. From (\ref{opers}), (\ref{H}), we have
\begin{eqnarray}
(\Psi, H\Psi)=\int\limits^{\infty}_{0}dx\left[\eta^*H_{11}\eta+\chi^*H_{22}\chi+V^*H_{33}V\right]+\nonumber\\
+\int\limits^{\infty}_{0}dx\left[\eta^*H_{13}V+\chi^*H_{23}V+V^*H_{31}\eta+V^*H_{32}\chi\right].
\nonumber
\end{eqnarray}
We estimate for the first integral is obtained by dropping the positive terms:
\begin{eqnarray}
\int\limits^{\infty}_{0}dx\left[\eta^*\left(-\frac{d^2}{dx^2}+1-\frac{2}{\cosh^2 x}\right)\eta+
\chi^*\left(-\frac{d^2}{dx^2}+1-\frac{2}{\cosh^2 x}\right)\chi\right.+\nonumber\\
+\left.V^*\left(-\frac{d^2}{dx^2}+\frac{8}{\kappa^2}-\left(\frac{8}{\kappa^2}+2\right)\frac{1}{\cosh^2 x}\right)V\right]=\nonumber\\
\int\limits^{\infty}_{0}dx\left[\eta^*\left(-\frac{d^2}{dx^2}+2\tanh^2x-1\right)\eta+
\chi^*\left(-\frac{d^2}{dx^2}+2\tanh^2x-1\right)\chi\right.+\nonumber\\
+\left.V^*\left(-\frac{d^2}{dx^2}+\left(\frac{8}{\kappa^2}+2\right)\tanh^2x-2\right)V\right]\ge\nonumber\\
\ge-\int\limits^{\infty}_{0}dx\left[\eta^*\eta+\chi^*\chi+2V^*V\right]\ge -2\int\limits^{\infty}_{0}dx\left[\eta^*\eta+\chi^*\chi+V^*V\right]
=-2\left(\Psi,\Psi\right)=-2.
\label{diag}
\end{eqnarray}
Taking into account that $\displaystyle\left|\frac{\tanh x}{\cosh x}\right|\le\frac{1}{2}$,
we obtain the estimate for the second integral
\begin{eqnarray}
\left|\int\limits^{\infty}_{0}dx\ 4\:\sqrt{\frac{1}{\kappa^2}-\frac{1}{2}}\:\frac{\tanh x}{\cosh x}\left[V^*(\chi+\eta)+(\chi+\eta)^*V\right]\right|\le\nonumber\\
\le4\:\sqrt{\frac{1}{\kappa^2}-\frac{1}{2}}\int\limits^{\infty}_{0}dx\left|\frac{\tanh x}{\cosh x}\left[V^*(\chi+\eta)+(\chi+\eta)^*V\right]\right|\le\nonumber\\
\le2\:\sqrt{\frac{1}{\kappa^2}-\frac{1}{2}}\int\limits^{\infty}_{0}dx\left|V^*(\chi+\eta)+(\chi+\eta)^*V\right|.
\label{ndiag1}
\end{eqnarray}
Further, applying the transformation
\begin{equation}
\eta=(u+w)/2-t/\sqrt{2}, \quad \chi=(u+w)/2+t/\sqrt{2}, \quad V=(u-w)/\sqrt{2}
\label{matr}
\end{equation}
and noting that it is unitary (does not change the value of the scalar product
$(\Psi,\Psi)$, we obtain
\begin{eqnarray}
\int\limits^{\infty}_{0}dx\left|V^*(\chi+\eta)+(\chi+\eta)^*V\right|=
\sqrt{2}\int\limits^{\infty}_{0}dx\left|u^*u-w^*w\right|\le\nonumber\\
\le\sqrt{2}\int\limits^{\infty}_{0}dx\left(u^*u+w^*w\right)\le
\sqrt{2}\int\limits^{\infty}_{0}dx\left(u^*u+w^*w+t^*t\right)=
\sqrt{2}\left(\Psi,\Psi\right)=\sqrt{2}.
\label{ndiag2}
\end{eqnarray}
From (\ref{ndiag1}) and (\ref{ndiag2}), we obtain
\begin{eqnarray}
\left|\int\limits^{\infty}_{0}dx\ 4\:\sqrt{\frac{1}{\kappa^2}-\frac{1}{2}}\frac{\tanh x}{\cosh x}
\left[V^*(\chi+\eta)+(\chi+\eta)^*V\right]\right|\le
2\:\sqrt{2/\kappa^2-1}.
\nonumber
\end{eqnarray}
From this inequality and (\ref{diag}), it follows that
\begin{eqnarray}
\mu_{min}\ge -2\left(1+\sqrt{2/\kappa^2-1}\right).
\label{mumin}
\end{eqnarray}
Using (\ref{ReImAbsLambda}) -- (\ref{ImLambda}) and (\ref{mumin}), we finally obtain
the following estimates:
\begin{equation}
|\lambda|\le\sqrt{-\mu_{min}}\le\sqrt{2\left(1+\sqrt{2/\kappa^2-1}\right)},
\label{estim2}
\end{equation}
\begin{equation}
|\mbox{Re}\:\lambda|\le 1, \quad
|\mbox{Im}\:\lambda|\le\sqrt{-\mu_{min}}\le\sqrt{2\left(1+\sqrt{2/\kappa^2-1}\right)}.
\label{estim3}
\end{equation}

\section{Numerical study of stability}
\setcounter{equation}{0}

\subsection{Refiniment of the localization domain for the eigenvalues of problem
(\ref{equ1}) -- (\ref{condint})}
Along with analytic estimates (\ref{estim2}), (\ref{estim3}), the localization domain
for the eigenvalues of problem (\ref{equ1}) -- (\ref{condint}) can be determined
numerically starting from the estimate
\begin{equation}
|\lambda|\le\sqrt{-\mu_{min}(\kappa)}.
\label{estim4}
\end{equation}
To find the numbers $\mu _{min}(\kappa)$, we consider an accompanying
eigenvalue problem for the operator $H$ (see (\ref{opers}), (\ref{H})), i.e.,
the $\mu$-parameterized ODE system
\begin{equation}
\mu I\Psi - H\Psi = 0, \quad 0\le x<\infty,
\label{4.1!}
\end{equation}
subject to the boundary conditions (\ref{cond0}) -- (\ref{condinf3}) and requirement
(\ref{condint}).

In order to solve this problem numerically, it should be reduced to a problem
defined on a finite interval. As noted in Section 3, the condition that the solutions
belong to the space $L_2[0,\infty)$ is equivalent to the requirement that the
values of the solutions for large $x$ belong to a three-dimensional subspace of
$\mbox{\bf C}^6$, which, up to exponentially small terms, is defined as
\begin{eqnarray}
\eta^{\prime}(x_\infty ) = -\sqrt {1 - \mu }\ \eta (x_\infty ), \label{4.2!}\\
\chi^{\prime}(x_\infty ) = -\sqrt {1 - \mu }\ \chi (x_\infty ), \label{4.3!}\\
V^{\prime}(x_\infty ) = -\sqrt {8/\kappa ^2 - \mu }\ V(x_\infty ) \label{4.4!},
\end{eqnarray}
where $\mu<1$ and $x_{\infty}\gg 1$  (cf. with (\ref{approx1}) -- (\ref{approx3})).

Problem (\ref{4.1!}), (\ref{cond0}), (\ref{4.2!})--(\ref{4.4!}) is self-adjoint. Taking into
account item 1 of Section 3, we immediately find that, for $\kappa = \sqrt {2}$,
$\mu = 0$ is its only eigenvalue (of multiplicity 3), and that the corresponding
eigenfunctions are given by (\ref{eig1}) -- (\ref{eig3}). For $\kappa$ different
from $\sqrt{2}$, the eigenvalue $\mu=0$ has multiplicity 2, with the corresponding
eigenfunctions being given by (\ref{eig5}) and (\ref{eig6}), and, as computations
show, the third zero eigenvalue starts to move to the region of negative values.
The  eigenvalue is sought numerically by transferring boundary conditions
(\ref{4.2!}) -- (\ref{4.4!}) from the point $x=x_{\infty}$ to the point $x=0$ and examining
the changes of sign of the resulting determinant. Note that, in this case, a variant
of the sweep method is used \cite{9} (with regard to the studies on the robust
use of orthogonal sweep methods in singular problems \cite{10}).

Our computations show that, when $\kappa $ varies from $\sqrt {2}$ to zero,
the double zero eigenvalue remains unchanged, whereas the single eigenvalue
(which is just the eigenvalue $\mu _{min}$ we are interested in) moves to the
region of negative values (Fig. 1). The comparison of these results with
(\ref{mumin}) shows that the bound obtained analytically is too high. For example,
for $\kappa = 1.2$, the value of $\sqrt {-\mu _{min}}$ found numerically is equal to
0.45, whereas (\ref{mumin}) yields 1.8. Nevertheless, for $\kappa\la 1.23$, the
magnitude of $\sqrt {-\mu _{min}}$ is greater than $\sqrt {2}-1$ (see estimate
(\ref{estim1}) and Fig. 2), which complicates the localization of the eigenvalues
of problem (\ref{equ1})--(\ref{condint}) (see below).

\begin{figure}[h]
\includegraphics*[width=15cm, height=15cm, keepaspectratio, angle=-90]{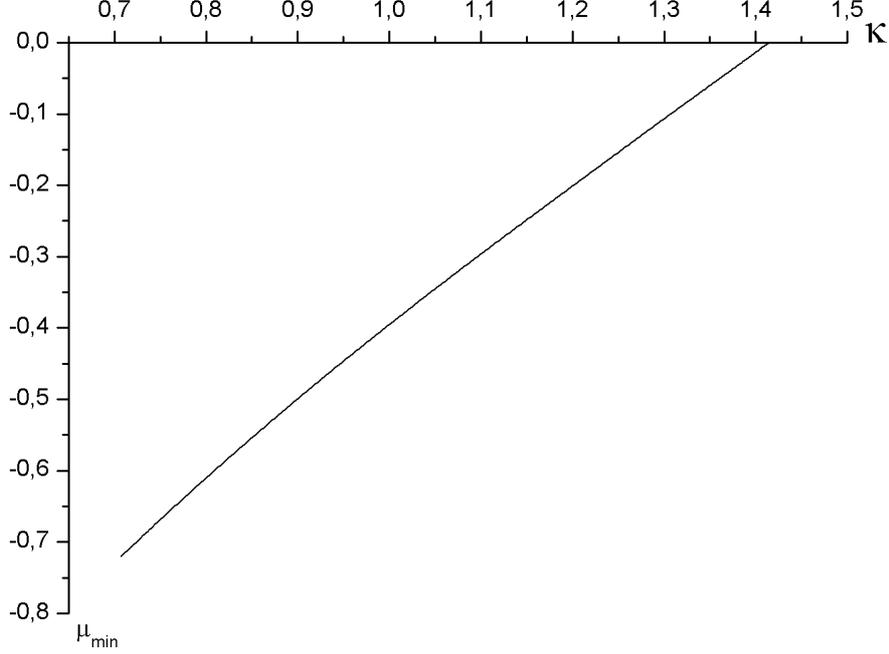}
\caption{the dependence of $\mu_{min}$ from $\kappa$.}
\label{fig1}
\end{figure}

\begin{figure}[h]
\includegraphics*[width=15cm, height=15cm, keepaspectratio, angle=-90]{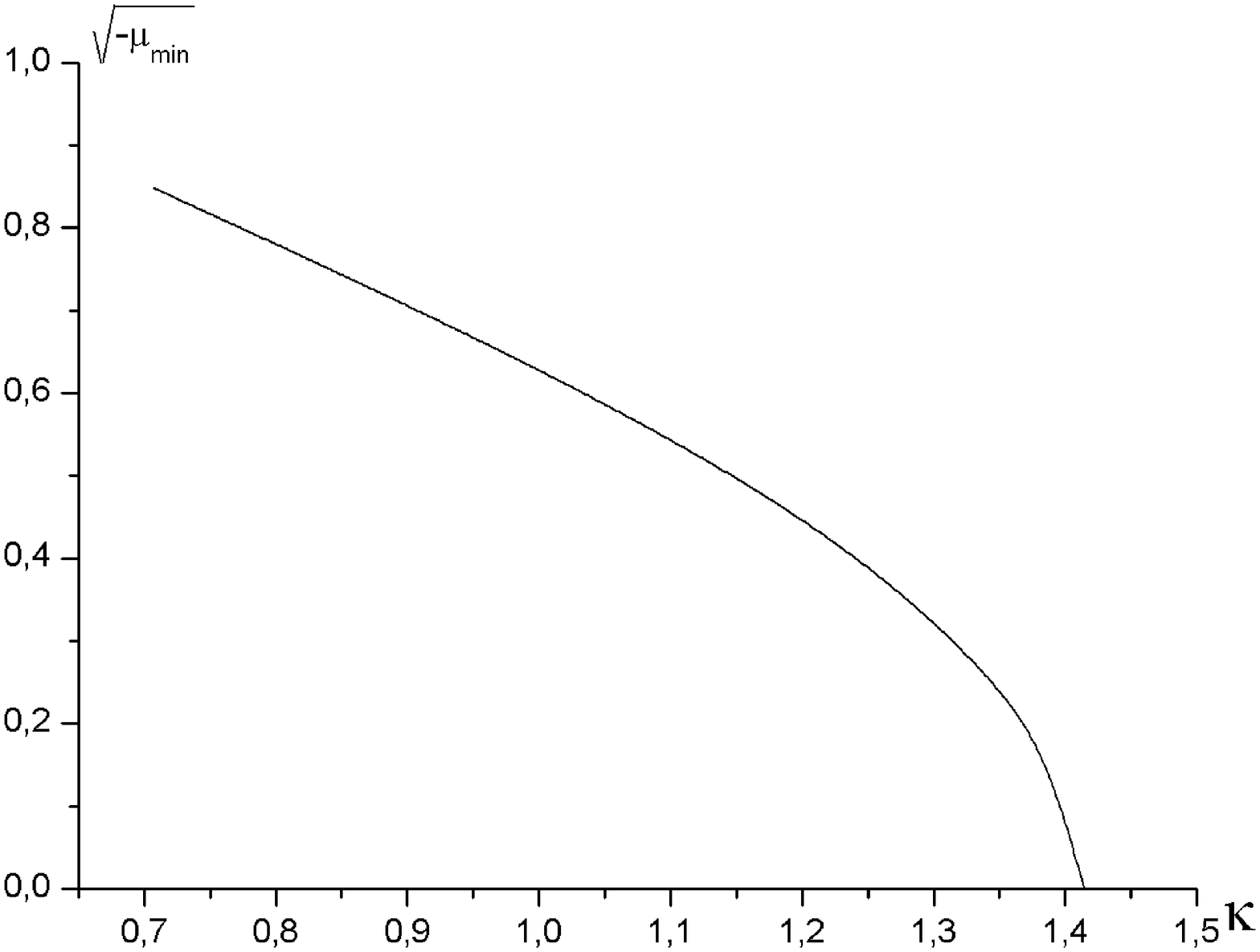}
\caption{the dependence of $\sqrt{-\mu_{min}}$ from $\kappa$.}
\label{fig2}
\end{figure}

\subsection{Search for eigenvalues of problem (\ref{equ1}) -- (\ref{condint})
in the sector (\ref{estim4}), (\ref{sector})}

Now, let us consider problem (\ref{equ1}) -- (\ref{condint}) itself. Its reduction
to a problem defined on a finite interval is similar to that for the previous
problem for the ODE (\ref{4.1!}). Representation (\ref{approx1}) -- (\ref{approx3})
yields the following boundary conditions at $x=x_{\infty}$:
\begin{equation}
\eta^{\prime}(x_\infty )=-\sqrt{1+2\lambda-\lambda^2}\:\eta(x_\infty),
\label{inf1}
\end{equation}
\begin{equation}
\chi^{\prime}(x_\infty)=-\sqrt{1-2\lambda-\lambda^2}\:\chi(x_\infty),
\label{inf2}
\end{equation}
\begin{equation}
V^{\prime}(x_\infty)=-\sqrt{8/\kappa^2-\lambda^2}\:V(x_\infty),
\label{inf3}
\end{equation}
where the roots are assumed to have positive real parts.

The nonsingular spectral problem (\ref{equ1}) -- (\ref{cond0}), (\ref{inf1}) -- (\ref{inf3})
was studied numerically by a method based on the generalization of the
argument principle (methods of localization of descrete spectrum points
based on the argument principle and its modifications are discussed, for
example, in \cite{7} -- \cite{88} and in references therein) with the use of the
differential sweep \cite{9} (as noted earlier, the problems of the robust
application of the method from \cite{9}, as well as other modifications
of the sweep method, to singular eigenvalue problems are discussed in
\cite{10}). Conditions (\ref{inf1}) -- (\ref{inf3}) are represented in the form
\begin{equation}
\varphi(x_{\infty},\lambda)\:W(x_{\infty},\lambda)=0,
\label{infall}
\end{equation}
where $W=(\eta,\eta^{\prime}, \chi, \chi^{\prime}, V, V^{\prime})^T$ and
\begin{equation}
\varphi (x_\infty ,\lambda )=
\left\|
\begin{array}{cccccc}
\sqrt {1+2\lambda - \lambda ^2}&1&0&0&0&0\\
0&0&\sqrt {1-2\lambda - \lambda ^2}&1&0&0\\
0&0&0&0&\sqrt {8/\kappa ^2 - \lambda ^2}&1\\
\end{array}
\right\|.
\label{varphi}
\end{equation}

As a result of the sweep, we obtain a condition at the point $x=0$ with
the $3\times 6$ matrix $\varphi(0,\lambda)$ equivalent to (\ref{infall}),
(\ref{varphi}). If we augment thi matrix up to a square one by adding
the lower matrix block corresponding to conditions (\ref{cond0}),
eigenvalue problem can be formulated as follows:

a number $\lambda$ is an eigenvalue $\Longleftrightarrow$ the determinant
of the resulting matrix is equal to zero;\\
the geometric multiplicity of an eigenvalue (i.e., the number of linear
independent eigenfunctions) is equal to the defect of the resulting
matrix.

Note that the coefficients of the matrix $\varphi(0,\lambda)$ and,
hence, the determinant of the resulting matrix, are not analytic
functions of $\lambda$. Nevertheless, as shown in \cite{77},
the argument principle can still be used for determining the number
of the eigenvalues in a given region of the spectral parameter $\lambda$.
The search can rely on the determinants $M_{245}$ for the first
condition and $M_{136}$ for the second condition in (\ref{cond0}),
respectively, where $M_{ijk}$ is the minor of the matrix $\varphi(0,\lambda)$
composed of the columns $i$, $j$, and $k$. The zeros of  det$(M_{ijk})$
correspond to the eigenvalues of problem (\ref{equ1}) -- (\ref{cond0}), (\ref{inf1}) -- (\ref{inf3}).
The algebraic multiplicity of an eigenvalue is the multiplicity of the
corresponding zero of the determinant.

From the computational results, it follows that when $\kappa$ is close to
$\sqrt {2}$, the search domain for unstable eigenvalues is a circle of
radius $\sqrt{-\mu _{min}}$ centered at the origin (see (\ref{estim4})).
When $\kappa$ approaches zero, the search domain is the same circle
with the cuts along the real axis from $\sqrt {2}-1$ to the right and from
$-\sqrt {2}+1$ to the left (see (\ref{estim1})).

The critical difficulty associated with this problem is that the points
lying on the cuts belong to the continuous spectrum (when  $\lambda$
is close to the cuts, the computation becomes unstable). Thus, the
methods of the localization of the eigenvalues described in \cite{7}--\cite{88}
cannot directly be applied to our problem, since the resulting determinant
(or some other function used in the argument principle) cannot
continuously be extended to the boundary of the domain. (Note
only that, if there were descrete eigenvalues appearing in the
intervals of the continuous spectrum, they would develop somehow
when $\kappa$ varies; however, the computation did not reveal
anything of this kind near the cuts when $\kappa$ varied.)

From the problem symmetry, it follows that the eigenvalues with
nonzero real and imaginary parts are located in the complex
$\lambda$-plane by quadruplets ($\lambda , -\lambda , \lambda^{*}, -\lambda^{*}$)
and those with only real or imaginary nonzero parts, by pairs on the
real or imaginary axis, respectively.

To study perturbations of the degenerate eigenvalue $\lambda = 0$
and reveal the presence / absence of complex eigenvalues,
we made computations on the following contours in the $\lambda$-plane:

circles of radius varying from $\varepsilon$ to $\sqrt{2}-1-\varepsilon$,
where $\varepsilon$ is a small number, centered at $\lambda=0$;

circular sectors of the angle  90$^0$ with the vertex at  $\lambda=i\:\varepsilon$
and radius $\sqrt{-\mu _{min}}$ located in the first quadrant (the eigenvalue
symmetry property is used);

intervals lying on the real or imaginary axis;

separate contours outside the real axis.

The computations show that, in the degenerate case of $\kappa=\sqrt{2}$,
the algebraic multiplicity of the eigenvalue $\lambda=0$ is equal to four,
and its geometric multiplicity is equal to three (cf. with (\ref{eig1}) -- (\ref{eig3})).
The difference in the two multiplicities is explained not by the presence of
Jordan block (which is impossible in view of the self-adjointness of the
boundary value problem) but by the fact that Eq. (\ref{equ3}) contains
the square of the spectral parameter $\lambda$.

When $\kappa$ varies from $\sqrt{2}$ to 0, the following picture is observed.
As before, the algebraic multiplicity of the eigenvalue $\lambda=0$ is equal
to 4. Thus, we may conclude that the eigenvalue $\lambda=0$ is not split
into several eigenvalues. Moreover, when $\kappa$ is close to $\sqrt{2}$,
it follows from the results obtained that the problem has no other eigenvalues.
When we move from $\kappa=\sqrt{2}$, the geometric multiplicity reduces
from 3 to 2 (instead of eigenfunctions (\ref{eig2}), (\ref{eig3}) we have (\ref{eig5})).
The reduction of the geometric multiplicity upon the perturbation of a parameter
is a typical phenomenon for quadratic pencils (in contrast to linear ones).
Eigenfunctions (\ref{eig1}) -- (\ref{eig3}) for $\kappa=\sqrt{2}$ and (\ref{eig5}), (\ref{eig6})
for other $\kappa$ are found numerically with accuracy up to $10^{-4}$.
As $\kappa$ approaches zero, problem (\ref{equ1}) -- (\ref{condint})
becomes more stiff. In the computations with fixed (double) relative accuracy,
the lowest value of $\kappa$ was as small as 0.05.

For additional stability control for the eigenvalue $\lambda=0$, we used
the following consideration. From the symmetry of the eigenvalues and the
existence of the eigenfunctions (\ref{eig5}), (\ref{eig6}), it follows that, at least,
the property of being repeated for the eigenvalue $\lambda=0$ is conserved.
Hence, the perturbation could accur only through splitting of two symmetric
eigenvalues along one of the axis. We have checked numerically that this
effect did not take place (with regard to estimates (\ref{estim1}) and (\ref{estim3})).
No other eigenvalues have been found in the numerical experiments.

It may be concluded (with the greatest certainty for the interval $\kappa$:
$1.23\la\kappa<\sqrt{2}$) that the singular boudary value problem,
(\ref{equ1}) -- (\ref{condint}) has only one eigenvalue $\lambda=0$
of algebraic multiplicity 4 and geometric multiplicity 2 with the
corresponding eigenfunctions (\ref{eig5}), (\ref{eig6}). No complex
eigenvalues have been found in the admissible domain (\ref{estim4}).

\section {Conclusions}

The analytical and numerical studies reported in this paper allow us
to conclude that solution (\ref{phisol}), (\ref{xisol}) is dynamically stable
with respect to small perturbations of form  (\ref{deltaxi}), (\ref{deltaphi})
for $0<\kappa<\sqrt{2}$. This statement has the greatest certainty in the
interval $\kappa$: $1.23\la\kappa<\sqrt{2}$. Moreover, solution (\ref{phisol}),
(\ref{xisol}) in the course of its evolution cannot split into small nonlocalized
oscillations (along the component $\xi$) over kink (\ref{wall}) since it is not
advantageous from the energy standpoint. The question of whether the
topological Q-ball (\ref{phisol}), (\ref{xisol}) is absolutely stable in the sector
$\{P_0, Q_0\}$ is still open.

\section {Acknowledgments}

This work was supported by the Russian Foundation for Basic Research,
projects 02-01-00050 and 00-15-96562. We also acknowledge the support
of the Federal Agency of Atomic Energy of the Russian Federation.

We are grateful to A. A. Abramov, N. A. Voronov, A. E. Kudryavtsev and
S. P. Popov for discussion of this work and usefull comments.

\end{document}